\documentclass[11pt,bib]{ba}

\usepackage{bayes}
\usepackage{amsmath,amsfonts,amsthm,amssymb,latexsym}
\usepackage[pdftex]{graphicx}
\usepackage{graphicx}
\renewcommand{\bibitem}{\vskip 1.45pt \par\hangindent\parindent\hskip-\parindent}

\begin{document}

\inserttype{article}

\title{Simulation-efficient~shortest~probability~intervals}
\author{Ying Liu, Andrew Gelman, and Tian Zheng}{Ying Liu, Andrew Gelman, and Tian Zheng\\Department of Statistics, Columbia University, New York\\[.2cm]29 Jan 2013}

\maketitle

Bayesian highest posterior density (HPD) intervals can be estimated
directly from simulations via empirical shortest intervals.
Unfortunately, these can be noisy (that is, have a high Monte Carlo
error). We derive an optimal weighting strategy using
bootstrap and quadratic programming to obtain a more computationally stable HPD, or in general, shortest probability interval (Spin). We prove the consistency of our method. Simulation studies on a range of theoretical and real-data examples, some with symmetric and some with asymmetric posterior densities, show that intervals constructed using Spin have better coverage (relative to the posterior distribution) and lower Monte Carlo error than empirical shortest intervals. We implement the new method in an R package ({\tt SPIn}) so it can be routinely used in post-processing of Bayesian simulations.

Key words:  Bayesian computation, highest posterior density, bootstrap.

\section{Introduction}
It is standard practice to summarize Bayesian inferences via posterior intervals of specified coverage (for example, 50\% and 95\%) for parameters and other quantities of interest.  In the modern era in which posterior distributions are computed via simulation, we most commonly see central intervals:  the $100(1\!-\!\alpha)\%$ central interval is defined by the $\frac{\alpha}{2}$ and $1\!-\!\frac{\alpha}{2}$ quantiles.  Highest-posterior density (HPD) intervals (recommended, for example, in the classic book of Box and Tiao, 1973) are easily determined for models with closed-form distributions such as the normal and gamma but are more difficult to compute from simulations.

\begin{figure}
\centering
\includegraphics[width=.8\textwidth]{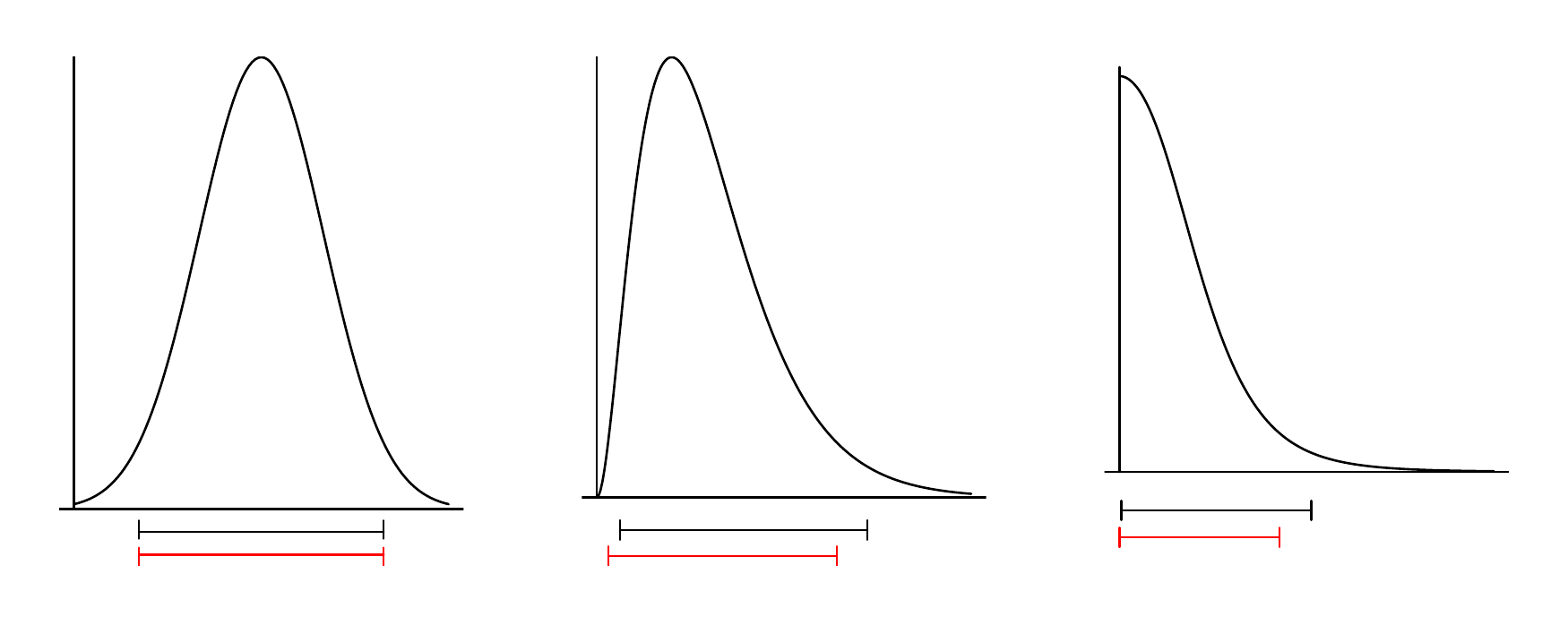}
\caption{\em Simple examples of central (black) and highest probability density (red) intervals.  The intervals coincide for a symmetric distribution; otherwise the HPD interval is shorter.  The three examples are a normal distribution, a gamma with shape parameter 3, and the marginal posterior density for a variance parameter in a hierarchical model.}
\label{fig1}
\end{figure}

We would like to go back to the HPD, solving whatever computational problems necessary to get it to work.  Why?  Because for an asymmetric distribution, the HPD interval can be a more reasonable summary than the central probability interval. Figure \ref{fig1} shows these two types of intervals for three distributions: for symmetric densities (as shown in the left panel in Figure \ref{fig1}), central and HPD intervals coincide; whereas for the two examples of asymmetric densities (the middle and right panels in Figure \ref{fig1}), HPDs are shorter than central intervals (in fact, the HPD is the shortest interval containing a specified probability).

In particular, when the highest density occurs at the boundary (the
right panel in Figure \ref{fig1}), we strongly prefer the shortest probability interval to the central interval; the HPD covers the highest density part of the distribution and also the mode. In such cases, central intervals can be much longer and have the awkward property at cutting off a narrow high-posterior slice that happens to be near the boundary, thus ruling out a part of the distribution that is actually strongly supported by the inference.

One concern with highest posterior density intervals is that they
depend on parameterization.  For example, the left endpoint of the HPD
in the right panel of Figure \ref{fig1} is 0, but the interval on the
logarithmic scale does not start at $-\infty$.  Interval estimation is always conditional on the purposes to which the estimate will be used.  Beyond this, univariate summaries cannot completely capture multivariate relationships.  Thus all this work is within the context of routine data analysis (e.g., Spiegelhalter et al., 1994, 2002) in which interval estimates are a useful way to summarize inferences about parameters and quantities of interest in a model in understandable parameterizations.  We do not attempt a conclusive justification of HPD intervals here; we merely note that in the pre-simulation era such intervals were considered the standard, which suggests to us that the current preference for central intervals arises from computational reasons as much as anything else.

\begin{figure}
\centering
\includegraphics[width=\textwidth]{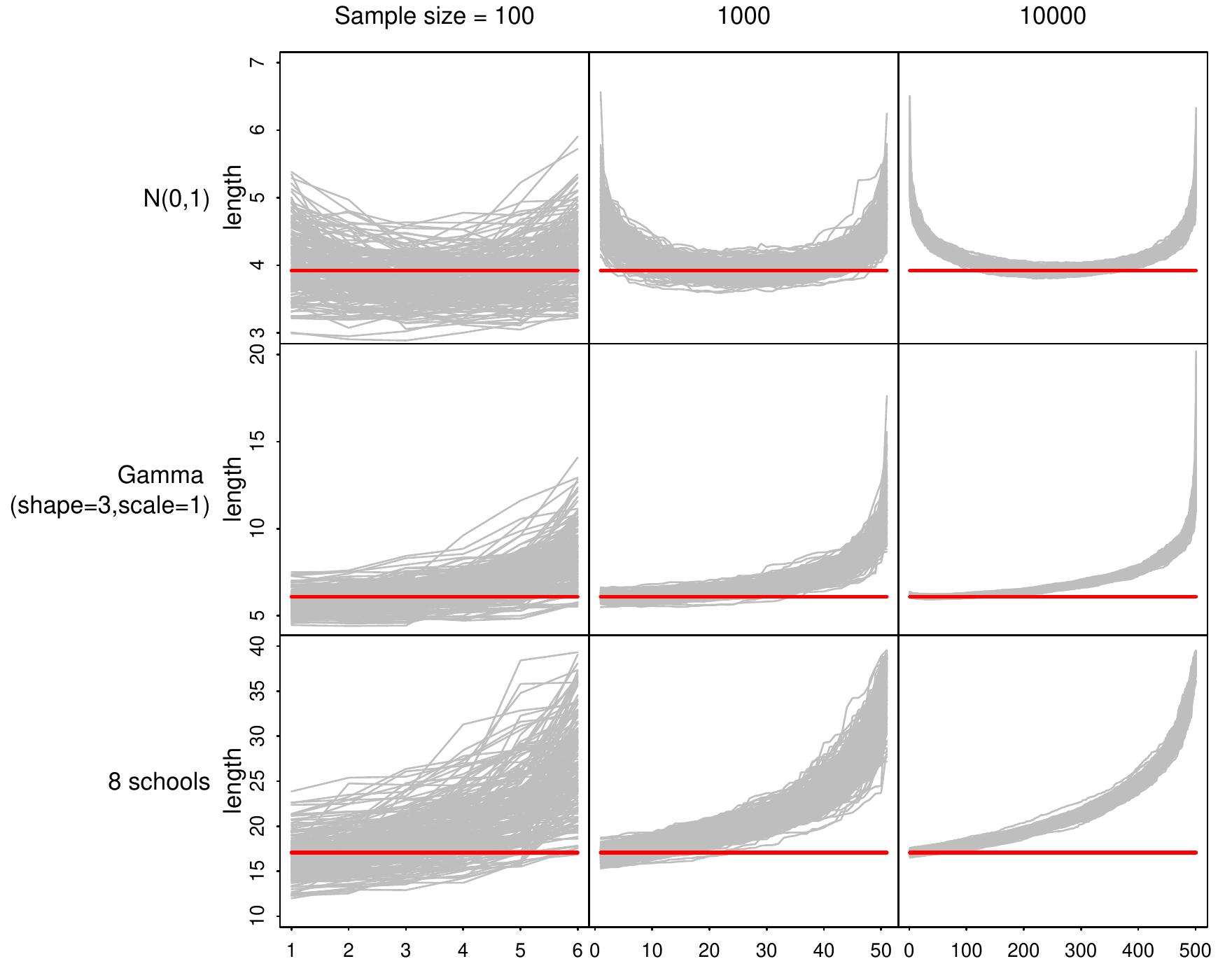}
\vspace{.1cm}
\caption{\em Lengths of 95\% empirical probability intervals from several simulations for each of three models.   Each gray curve shows interval length as a function of the order statistic of the interval's lower endpoint; thus, the minimum value of the curve corresponds to the empirical shortest 95\% interval.  For the (symmetric) normal example, the empirical shortest interval is typically close to the central interval (for example, with a sample of size 1000, it is typically near $(x_{(26)},x_{(975)})$).  The gamma and eight-schools examples are skewed with a peak near the left of the distribution, hence the empirical shortest intervals are typically at the left end of the scale.  The red lines show the lengths of the true shortest 95\% probability interval for each distribution.  The empirical shortest interval approaches the true value as the number of simulation draws increases but is noisy when the number of simulation draws is small, hence motivating a more elaborate estimator.}
\label{fig2}
\end{figure}

For the goal of computing an HPD interval from posterior simulations, the most direct approach is the {\em empirical shortest probability interval}, the shortest interval of specified probability coverage based on the simulations
(Chen and Shao, 1999).  For example, to obtain a 95\% interval from a
posterior sample of size $n$, you can order the simulation draws and
then take the shortest interval that contains $0.95n$ of the draws.
This procedure is easy, fast, and simulation-consistent (that is, as
$n\!\rightarrow\!\infty$ it converges to the actual HPD interval
assuming that the HPD interval exists and is unique). The only trouble
with the empirical shortest probability interval is that it can be too
noisy, with a high Monte Carlo error (compared to the central
probability interval) when computed from the equivalent of a small number of simulation
draws.  This is a concern with current Bayesian methods that rely on
Markov chain Monte Carlo (MCMC) techniques, where for some problems the effective sample size of the posterior draws can be low (for example, hundreds of thousands of steps might be needed to obtain an effective sample size of 500).

Figure \ref{fig2} shows the lengths of the empirical shortest 95\%
intervals based on several simulations for the three distributions shown in Figure \ref{fig1}, starting from the $k$th order statistic. For each distribution and each specified number of independent simulation draws, we carried out 200 replications to get a sense of the typical size of the Monte Carlo error. The lengths of the 95\% intervals are highly variable when the number of simulation draws is small.

\begin{figure}
\centering
\includegraphics[width=\textwidth]{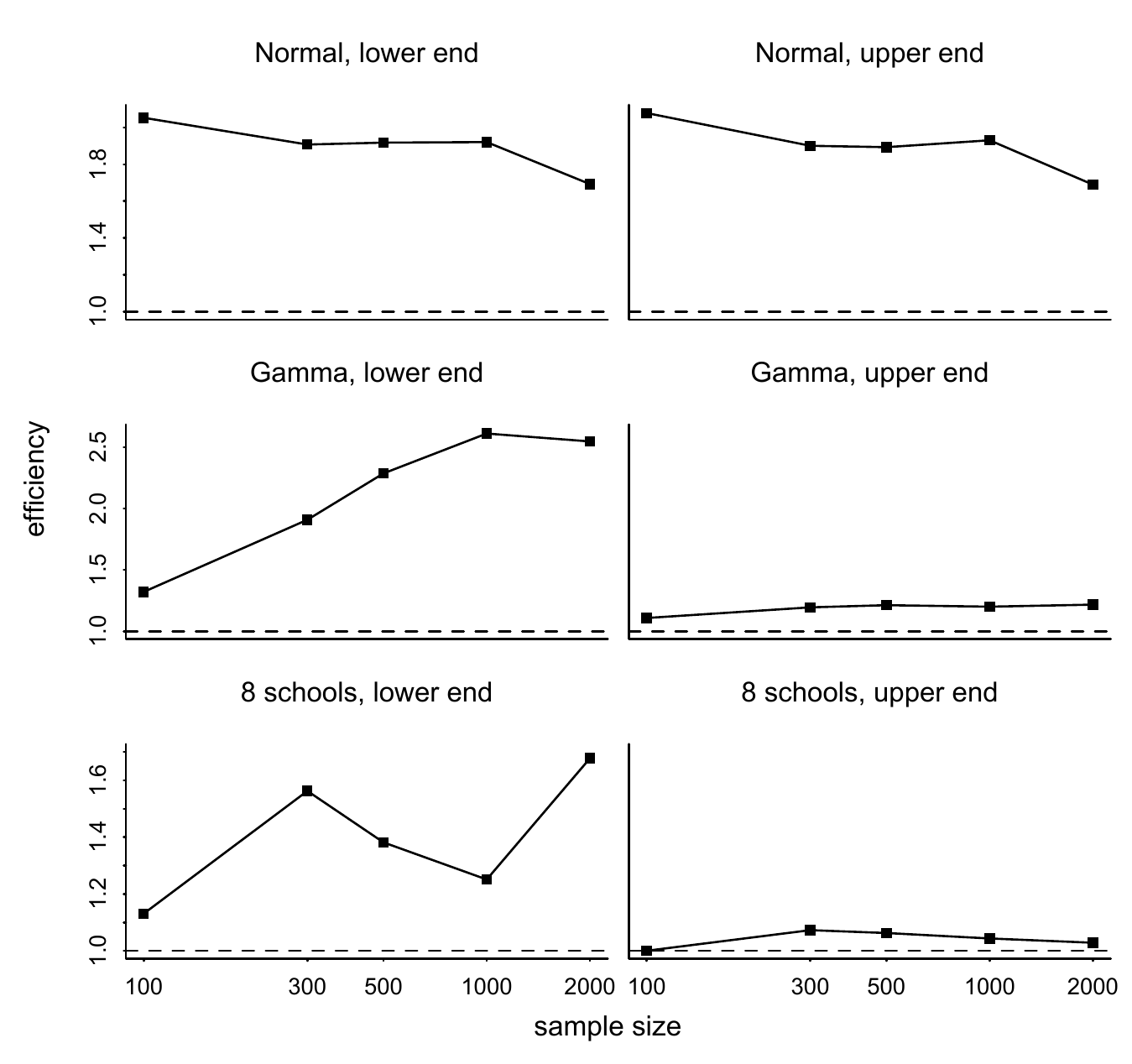}
\vspace{.1cm}
\caption{\em Efficiency of Spin for 95\% shortest intervals for the three distributions shown in
  Figure \ref{fig1}. For the
  eight-schools example, Spin is compared to a modified empirical HPD
  that includes the zero point in the simulations. The efficiency is always greater than 1, indicating that Spin always outperforms the empirical HPD.  The jagged appearance of some of the lines may arise from discreteness in the order statistics for the 95\% interval.}
\label{eff}
\end{figure}

In this paper, we develop a quadratic programming strategy coupled
with bootstrapping to estimate the endpoints of the shortest
probability interval. Simulation studies show that our procedure,
which we call Spin, results in more stable endpoint estimates compared
to the empirical shortest interval (Figure \ref{eff}). Specifically,
define the efficiency as
\begin{equation}
\mbox{efficiency} = \frac{\mbox{MSE}(\mbox{empirical shortest interval})}{\mbox{MSE}(\mbox{Spin})},\nonumber
\end{equation}
so that an efficiency greater than 1 means that Spin is more efficient. We show
in Figure \ref{eff} that, in all cases that we
experimented on, Spin is more efficient than the
competition. We derive our method in Section \ref{methods}, apply it to some theoretical examples in Section \ref{res} and in two real-data Bayesian analysis problems in Section \ref{res2}. We have
implemented our algorithm as  {\tt SPIn}, a publicly available package
in R (R Development Core Team, 2011).

\section{Methods}\label{methods}

\subsection{Problem setup}

Let $X_1$,\dots, $X_n\stackrel{iid}{\sim}F$, where $F$ is a continuous unimodal
distribution. The goal is to estimate the $100(1-\alpha)\%$ shortest
probability interval for $F$. Denote the true shortest probability interval by $(l(\alpha),u(\alpha))$. Define $G=1-F$, so that
$F(l(\alpha))+G(u(\alpha))=\alpha$.

\begin{figure}
\centering
\vspace{-1.5cm}
\includegraphics[width=140mm]{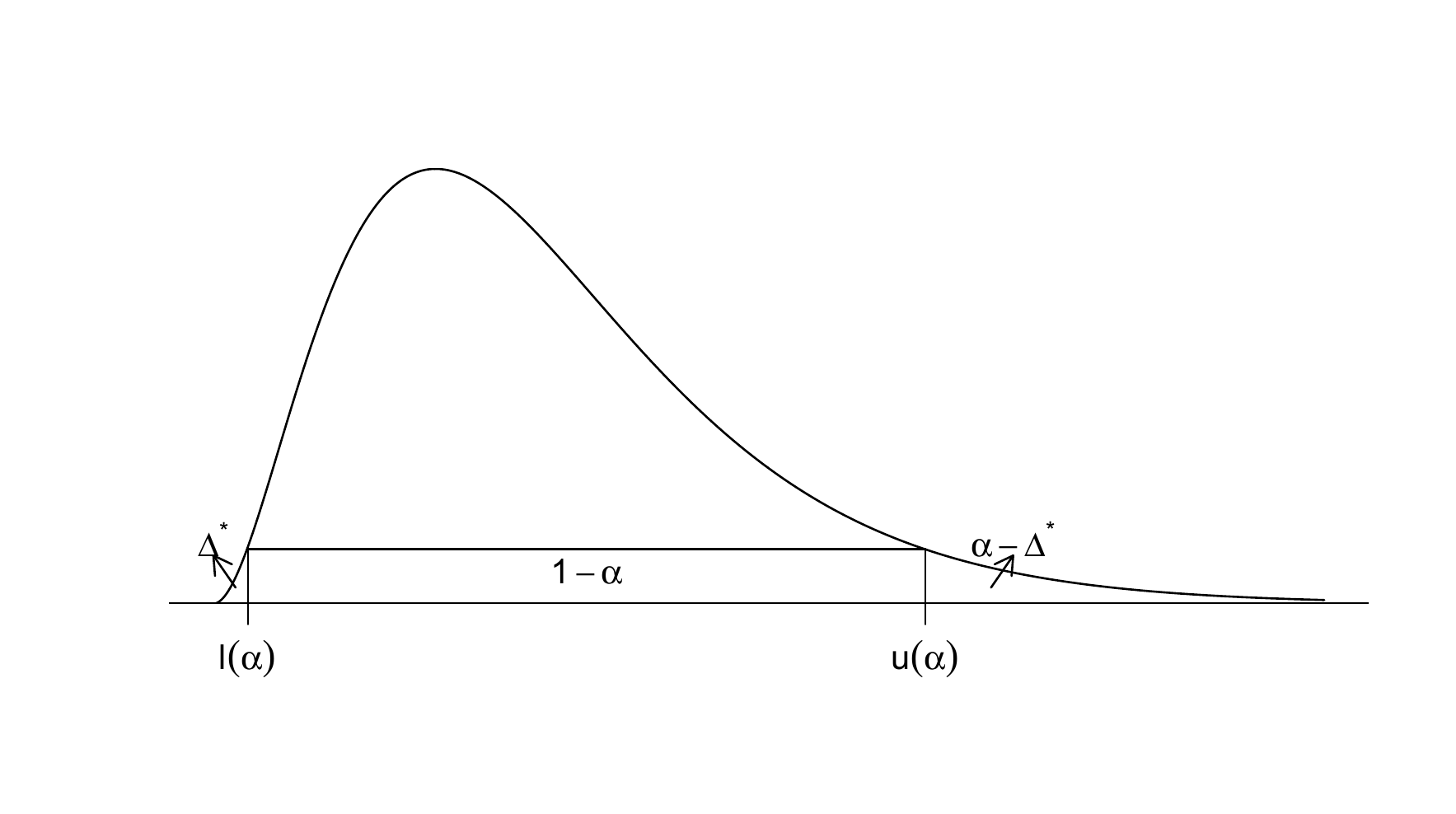}
\vspace{-1.2cm}
\caption{\em Notation for shortest probability intervals.}
\label{def}
\end{figure}

To estimate the interval, for $0\le\Delta\le\alpha$, find $\Delta$ such that
$G^{-1}(\alpha-\Delta)-F^{-1}(\Delta)$
is a minimum, i.e.,
\begin{equation}
\Delta^*=\mbox{argmin}_{\Delta\in[0,\alpha]}\{G^{-1}(\alpha-\Delta)-F^{-1}(\Delta)\}.\nonumber
\end{equation}
Taking the derivative,
\begin{eqnarray}
\frac{\partial}{\partial\Delta}[(1-F)^{-1}(\alpha-\Delta)-F^{-1}(\Delta)]=0,\nonumber
\end{eqnarray}
we get
\begin{equation}\label{d}
\frac{1}{f(G^{-1}(\alpha-\Delta))}-\frac{1}{f(F^{-1}(\Delta))}=0,
\end{equation}
where $f$ is the probability density function of $X$. The minimum can only be attained at solutions to (\ref{d}), or $\Delta=0$ or
$\alpha$ (Figure \ref{def}). It can easily be shown that if $f'(x) \ne 0$ a.e., the solution to (\ref{d}) exists and is unique. Then
\begin{eqnarray}
l(\alpha)&=&F^{-1}(\Delta^*),\nonumber\\
u(\alpha)&=&G^{-1}(\alpha-\Delta^*).\nonumber
\end{eqnarray}

Taking the lower end for example, we are interested in a weighting strategy such that
$\hat{l} = {\sum}_{i=1}^{n}w_iX_{(i)}$ (where $\sum w_i=1$) has the minimum mean
squared error (MSE),
$\mathrm{E}\left(\left|\left|{\sum}_{i=1}^{n}w_iX_{(i)}-l(\alpha)\right|\right|^2\right)$.
It can also be helpful to calculate $\mathrm{MSE}(X_{([n\Delta^*])})=
\mathrm{E}\left(||X_{([n\Delta^*])}-l(\alpha)||^2\right)$.
In practice we estimate $\Delta^*$ by $\hat{\Delta}$ such that
\begin{eqnarray}\label{es}
\hat{\Delta}=\mbox{argmin}_{\Delta\in[0,\alpha]}\{\hat{G}^{-1}(\alpha-\Delta)-\hat{F}^{-1}(\Delta)\},
\end{eqnarray}
where $\hat{F}$ represents the empirical distribution and $\hat{G}=1-\hat{F}$. This yields the
widely used empirical shortest interval, which can have a high Monte Carlo error
(as illustrated in Figure \ref{fig2}). We will denote its endpoints by $l^*$
and $u^*$. The corresponding MSE for the lower endpoint is
$\mathrm{E}(||X_{([n\hat{\Delta}])}-l(\alpha)||^2)$.

\subsection{Quadratic programming}

Let $\hat{l}=\sum_{i=1}^{n}w_iX_{(i)}$.  Then
\begin{eqnarray}
\mathrm{MSE} (\hat{l}) & = & \mathrm{E}(\hat{l}-F^{-1}(\Delta^*))^2\nonumber\\
& = & \mathrm{E}\,(\hat{l}-\mathrm{E}\,\hat{l}+\mathrm{E}\,\hat{l}-F^{-1}(\Delta^*))^2\nonumber\\
& = &
\mathrm{E}\,(\hat{l}-\mathrm{E}\,\hat{l})^2+(\mathrm{E}\,\hat{l}-F^{-1}(\Delta^*))^2\nonumber\\
& = & \mathrm{Var}+\mathrm{Bias}^2,\nonumber
\end{eqnarray}
where
$\mathrm{E}(\hat{l})=\sum_{i=1}^{n}w_i\mathrm{E}X_{(i)}$
and $\mathrm{Var}=\sum_{i=1}^{n}w_i^2\mathrm{Var}X_{(i)}+2\sum_{i<j}w_iw_j\mathrm{cov}(X_{(i)},X_{(j)})$.
It has been shown (e.g., David and Nagaraja, 2003) that
\begin{eqnarray}
\mathrm{E}(X_{(i)})=Q_i+\frac{p_iq_i}{2(n+2)}Q_i''+o(n^{-1}),\nonumber
\end{eqnarray}
where $q_i=1-p_i$, $Q=F^{-1}$ is the quantile function, $Q_i=Q(p_i)=Q(\mathrm{E}U_{(i)})=Q(\frac{i}{n+1})$, and $Q_i''=\frac{Q_i}{f^2(Q_i)}$. Thus
\begin{eqnarray}
\mathrm{E}(\hat{l})\doteq\sum_{i=1}^{n}w_i\left(Q_i+\frac{p_iq_i}{2(n+2)}Q_i''\right). \label{eq_mean}
\end{eqnarray}
It has also been shown (e.g., David and Nagaraja, 2003) that
\begin{eqnarray}
\mathrm{Var}\,X_{(i)}&=&\frac{p_iq_i}{n+2}Q_i'^2+o(n^{-1})\nonumber\\
\mathrm{cov}(X_{(i)},X_{(j)}) & = &
\frac{p_iq_j}{n+2}Q_i'Q_j'+o(n^{-1}), \mbox{ for } i < j,\nonumber
\end{eqnarray}
where $Q_i'=\frac{1}{dp_i/dQ_i}=\frac{1}{f(Q_i)}$ ($f(Q_i)$ is called the density-quantile function). Thus,
\begin{eqnarray}
\mathrm{Var}(\hat{l})=\sum_{i=1}^{n}w_i^2\frac{p_iq_i}{n+2}Q_i'^2+2\sum_{i<j}w_iw_j\frac{p_iq_j}{n+2}Q_i'Q_j'+o(n^{-1}). \label{eq_var}
\end{eqnarray}
Putting \eqref{eq_mean} and \eqref{eq_var} together yields,
\begin{eqnarray}\label{MSE}
\mathrm{MSE}(\hat{l})&=&\sum_{i=1}^{n}w_i^2\frac{p_iq_i}{n+2}Q_i'^2+2\sum_{i<j}w_iw_j\frac{p_iq_j}{n+2}Q_i'Q_j' + \nonumber\\
&&+\,\left[\sum_{i=1}^{n}w_i(Q_i+\frac{p_iq_i}{2(n+2)}Q_i'')-Q(\Delta^*)\right]^2+o(n^{-1}).
\end{eqnarray}
Finding the optimal weights that minimize MSE as defined in \eqref{MSE} is then approximately a quadratic programming problem.

In this study we impose triangle kernels centered
around the endpoints of the empirical shortest interval on the weights for computational
stability. Specifically, the estimate of the lower endpoint has the
form,
\begin{eqnarray}
\hat{l}={\sum}_{i=i^*-b/2}^{i^*+b/2}w_iX_{(i)},\nonumber
\end{eqnarray}
where $i^*$ is the index of the endpoint of the empirical shortest
interval, $b$ is the bandwidth in terms of data points, and $w_i$
decreases linearly when $i$ moves away from $i^*$. We choose
$b$ to be of order $\sqrt{n}$ in this study. This optimization problem is equivalent to minimizing MSE with the following constraints:
\begin{eqnarray}\label{c}
\sum_{i=i^*-b/2}^{i^*+b/2}w_i&=&1\nonumber\\
\frac{w_{i}-w_{i-1}}{X_{(i)}-X_{(i-1)}}&=&\frac{w_{i-1}-w_{i-2}}{X_{(i-1)}-X_{(i-2)}}
\mbox{ for }i=i^*\!-b/2\!+\!2,\dots,i^*,i^*\!+\!2,\dots, i^*\!+\!b/2\nonumber\\
\frac{w_{i^*}-w_{i^*-1}}{X_{(i^*)}-X_{(i^*-1)}}&=&\frac{w_{i^*}-w_{i^*+1}}{X_{(i^*+1)}-X_{(i^*)}}\nonumber\\
w_{i^*-b/2}&\ge&0\nonumber\\
w_{i^*+b/2}&\ge& 0\nonumber\\
w_{i^*}-w_{i^*+1}&\ge&0.
\end{eqnarray}
The above constraints reflect the piecewise linear and symmetric pattern of the kernel. In practice, $Q$, $f$, and $\Delta^*$ can be substituted by the corresponding sample estimates $\hat{Q}$, $\hat{f}$, and $\hat{\Delta}$.

The above quadratic programming problem can be rewritten in the conventional matrix form,
\begin{eqnarray}
\mathrm{MSE}(\hat{l})\doteq\frac{1}{2}w^T\mathbf{D}w-d^Tw,\nonumber
\end{eqnarray}
where
\begin{eqnarray}
w=(w_{i^*-b/2},\dots,w_{i^*+b/2})^T,\nonumber
\end{eqnarray}
and $\mathbf{D}=(d_{ij})$ is a symmetric matrix with
\begin{eqnarray}
d_{ij}=\left\{\!\!
\begin{array}{rl}
 2(Q_i^2+\frac{p_iq_i}{n+2}Q_i'^2), & i=j\\
 2(\frac{Q_i'Q_j'}{n+2}p_iq_j+Q_iQ_j), & i<j,
\end{array} \right.\nonumber
\end{eqnarray}
\begin{eqnarray}
d^T=2Q(\Delta^*)Q_i,\nonumber
\end{eqnarray}
subject to
\begin{eqnarray}
\mathbf{A}^Tw \ge w_0,\nonumber
\end{eqnarray}
with appropriate $\mathbf{A}$ and $w_0$ derived from the linear
constraints in (\ref{c}).

\subsection{Proof of simulation-consistency of the estimated HPD}

The following result ensures the simulation-consistency of our endpoint
estimators when we use the empirical distribution and kernel density
estimate.

Under regularity conditions, with probability 1,
\begin{eqnarray}
\lim_{n \to \infty}\min_{w}\left(\frac{1}{2}w^T\hat{\mathbf{D}}_nw-\hat{d}_n^Tw\right) = \min_{w}\left(\frac{1}{2}w^T\mathbf{D}w-d^Tw\right),\nonumber
\end{eqnarray}
where $\hat{\mathbf{D}}_n$ and $\hat{d}_n$ are empirical estimates of $\mathbf{D}$ and $d$
based on empirical distribution function and kernel density estimates.

To see this, we first show that $\hat{\mathbf{D}}_n \to \mathbf{D}$ and
$\hat{d}_n \to d$ uniformly as $n \to \infty$ almost surely. By the
Glivenko-Cantelli theorem, $||\hat{F} - F||_\infty \stackrel{a.s.} \to 0$, which implies $\hat{Q} \leadsto Q$ almost surely,
where $\leadsto$ denotes weak convergence, i.e., $\hat{Q}(t) \to Q(t)$
at every $t$ where $Q$ is continuous (e.g., van der Vaart, 1998). It has
also been shown that $\int $E$_f(\hat{f}(x)\!-\!f(x))^2dx = O(n^{-4/5})$
under regularity conditions (see, e.g., van der Vaart, 1998), which
implies that $\hat{f}(x) \!\to\! f(x)$ almost surely for all $x$. The
endpoints of the empirical shortest interval are simulation-consistent (Chen and Shao, 1999).

The elements in matrix $\hat{\mathbf{D}}_n$ result from simple
arithmetic manipulations of $\hat{Q}$ and $\hat{f}$, so
$\hat{d}_{ij} \to d_{ij}$ with probability 1, which implies,
\begin{eqnarray}
\hat{\mathbf{D}}_n \to \mathbf{D} \mbox{ uniformly and almost surely,}\nonumber
\end{eqnarray}
given $\mathbf{D}$ is of finite dimension. We can prove the almost sure uniform convergence of
$\hat{d}_n$ to $d$ in a similar manner.

The optimization problem
$\min_{w}(\frac{1}{2}w^T\hat{\mathbf{D}}_nw-\hat{d}_n^Tw)$
corresponds to calculating the smallest eigenvalue of an augmented
matrix constructed from $\hat{\mathbf{D}}_n$ and $\hat{d}_n$. The above uniform
convergence then implies,
\begin{eqnarray}
\lim_{n \to \infty}\min_{w}(w^T\hat{\mathbf{D}}_nw-\hat{d}_n^Tw) = \min_{w}(w^T\mathbf{D}w-d^Tw).\nonumber
\end{eqnarray}

The same proof works for the upper endpoint.

\subsection{Bootstrapping the procedure to get a smoother estimate}

\begin{figure}
\centering
\includegraphics[width=\textwidth]{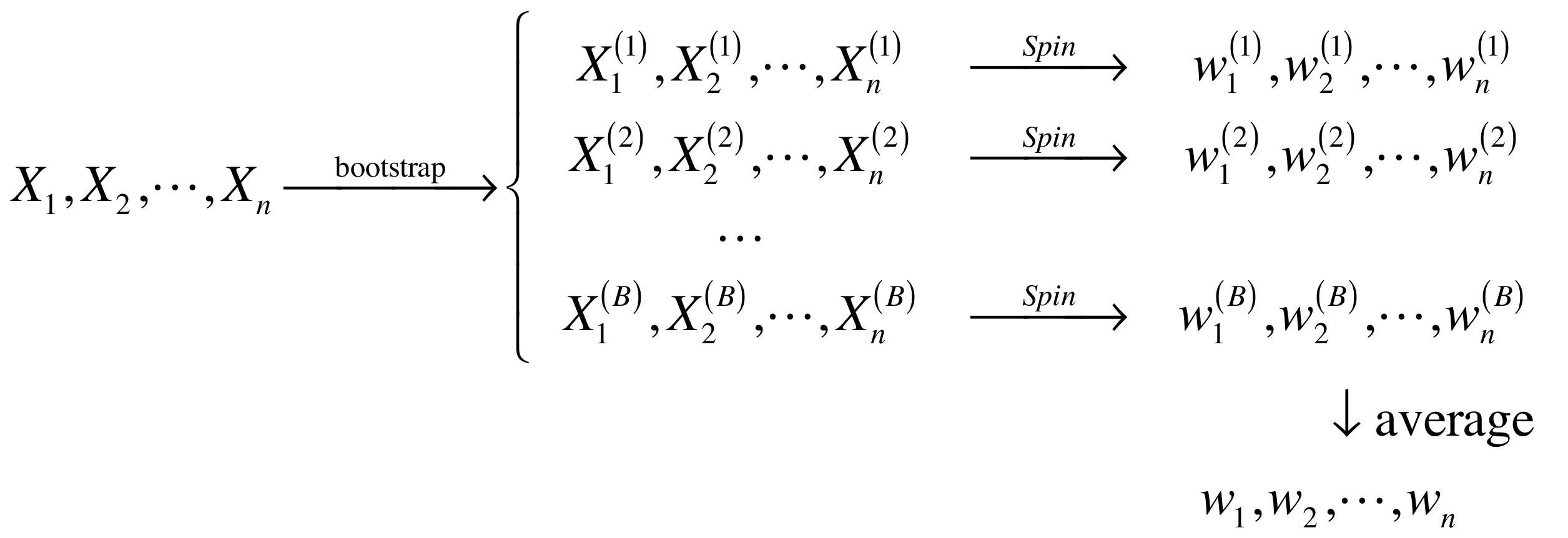}
\vspace{.2cm}
\caption{\em Bootstrapping procedure to get more stable weights.}
\label{bs}
\end{figure}

Results from quadratic programming as described above show that, as expected, Spin
has a much reduced bias than the empirical shortest
intervals. This is because the above procedure takes the shape of the
empirical distribution into account. However, the variance remains at
the same magnitude as that of the empirical shortest interval (as we shall see in the left
panel in Figure \ref{bvgamma}), because the optimal weights
derived from the empirical distribution are also subject to the same
level of variability as the empirical shortest intervals. We can use the bootstrap (Efron, 1979) to smooth away some of this noise and thus further reduce the
variance in the interval. Specifically, we bootstrap the original posterior draws $B$ times (in this study we set $B\!=\!50$) and calculate the Spin
optimal weights for each of the bootstrapped
samples. Here, we treat
the weights as general functions of the posterior distribution
under study rather than the endpoints of HPD interval of the posterior samples. We then compute the final
weights as the average of the $B$ sets of weights
obtained from the above procedure (Figure \ref{bs}).

\subsection{Bounded distributions}
As defined so far, our procedure necessarily yields an interval
within the range of the simulations.  This is undesirable if the
distribution is bounded with the boundary included in the HPD
interval (as in the right graph in Figure 1).  To allow boundary estimates, we augment our simulations with a
pseudo-datapoint (or two, if the distribution is known to be bounded on both sides).  For example, if a distribution is defined on $(0,\infty)$ then we insert another datapoint at 0; if the probability space is $(0,1)$, we insert additional points at 0 and 1.

\subsection{Discrete and multimodal distributions}
If a distribution is continuous and unimodal, the highest posterior
density region and shortest probability interval coincide.  More
generally, the highest posterior density region can be formed from
disjoint intervals. For distributions with known boundary of disjoint
parts, Spin can be applied to different regions separately and a HPD
region can be assembled using the derived disjoint intervals. When the
nature of the underlying true distribution is unknown and the sample
size is small, the inference of unimodality can be
difficult. Therefore, in this paper, we have focused on estimating the
shortest probability interval, recognizing that, as with interval estimates in general, our procedure is less relevant for multimodal distributions.

\section{Results for simple theoretical examples}\label{res}

We conduct simulation studies to evaluate the performance of our
methods. We generate independent samples from the normal, t(5), and gamma(3) distributions and construct 95\% intervals using these samples. We consider sample sizes of
100, 300, 500, 1000 and 2000. For each setup, we generate 20,000 independent 
replicates and use these to compute root mean squared errors (RMSEs) for upper and lower endpoints. We also construct
empirical shortest intervals as defined in (\ref{es}), parametric intervals and central intervals for
comparison. For parametric intervals, we calculate the sample mean
and standard deviation. For the normal distribution, the interval
takes the form of $\mbox{mean}\pm1.96\,\mbox{sd}$ (for the $t$ distribution we also
implement the same form as ``Gaussian approximation'' for comparison); for the gamma, we use the
mean and standard deviation to estimate its parameters first, and
then numerically obtain the HPD interval using the resulted density with
the two estimates plugged in. The empirical 95\%
central interval is defined as the 2.5\%th and 97.5\%th
percentiles of the sampled data points. We also use our methods to
construct optimal central intervals (see Section~\ref{discussion})
for the two symmetric distributions.

\begin{figure}
\centering
\includegraphics[width=150mm]{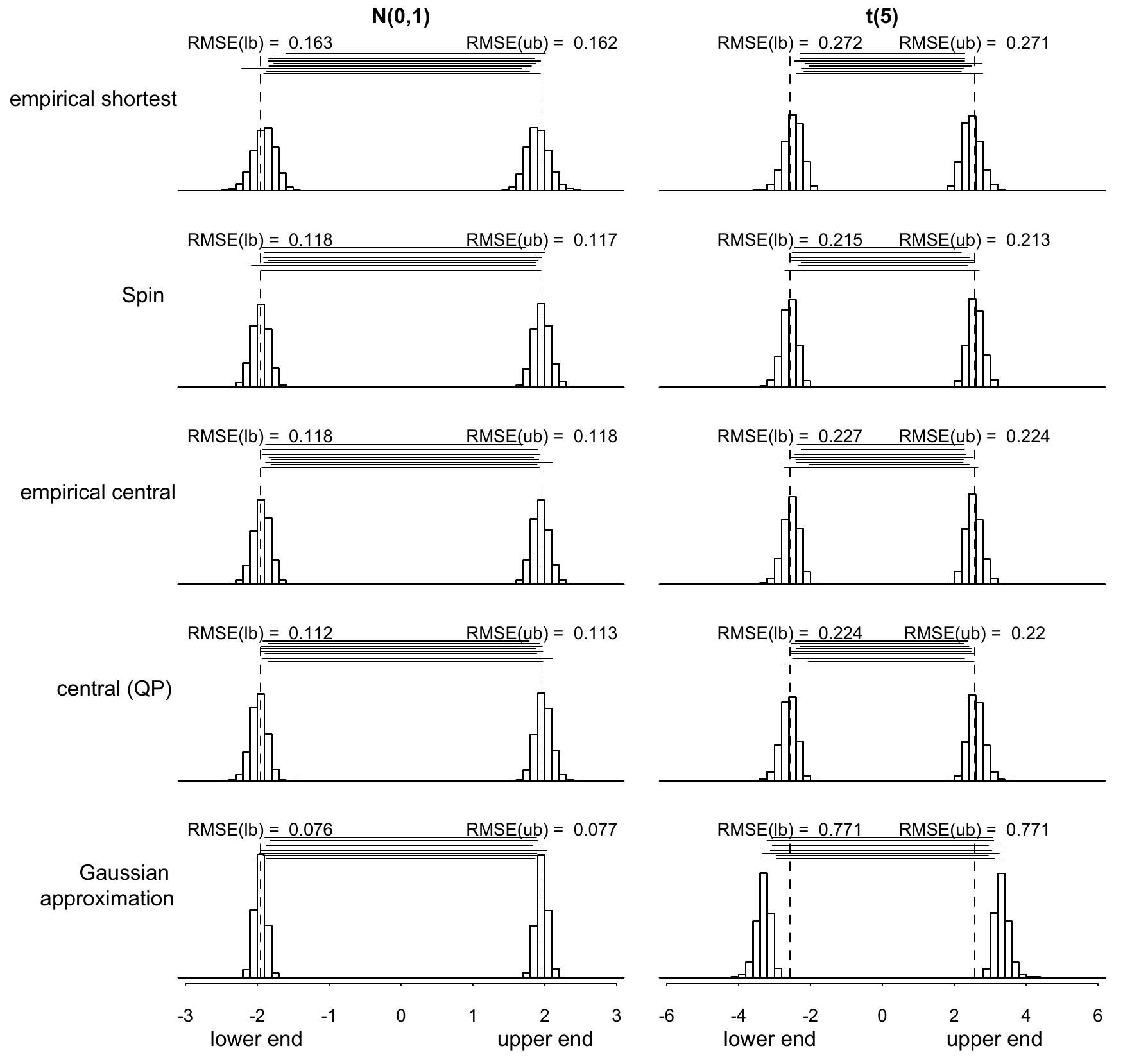}
\vspace{.2cm}
\caption{\em Spin for symmetric distributions: 95\% intervals for the normal and t(5) distributions, in each case based on 500 independent draws. Each horizontal bar
  represents an interval from one simulation. The histograms of the
  lower ends and the upper ends are based
  on results from 20,000 simulations. The dotted vertical lines
  represent the true endpoints of the HPD intervals.  Spin greatly outperforms the raw empirical shortest interval.  The central interval (and its quadratic programming improvement) does even better for the Gaussian but is worse for the t(5) and in any case does not generalize to asymmetric distributions.  The intervals estimated by fitting a Gaussian distribution do the best for the normal model but are disastrous when the model is wrong.}
\label{sym}
\end{figure}

Figure \ref{sym} shows the intervals constructed for the standard
normal distribution and the t(5) distribution 
based on 500 simulation draws. The empirical
shortest intervals tend to be too short in both cases, while Spins
have better endpoint estimates. Empirical central intervals are more stable than
empirical shortest intervals, and Spins have comparable RMSE for $\mbox{N}(0,1)$
and smaller RMSE for t(5). Our methods can further improve RMSE based on
the empirical central intervals as shown in the ``central (QP)'' row in
Figure \ref{sym}. The RMSE is the smallest if one specifies the correct parametric distribution
and uses that information to construct interval estimates, while in practice
the underlying distribution is usually not totally known, and mis-specifying it can result in far-off intervals (the right bottom panel in
Figure \ref{sym}).

\begin{figure}
\centering
\includegraphics[width=\textwidth]{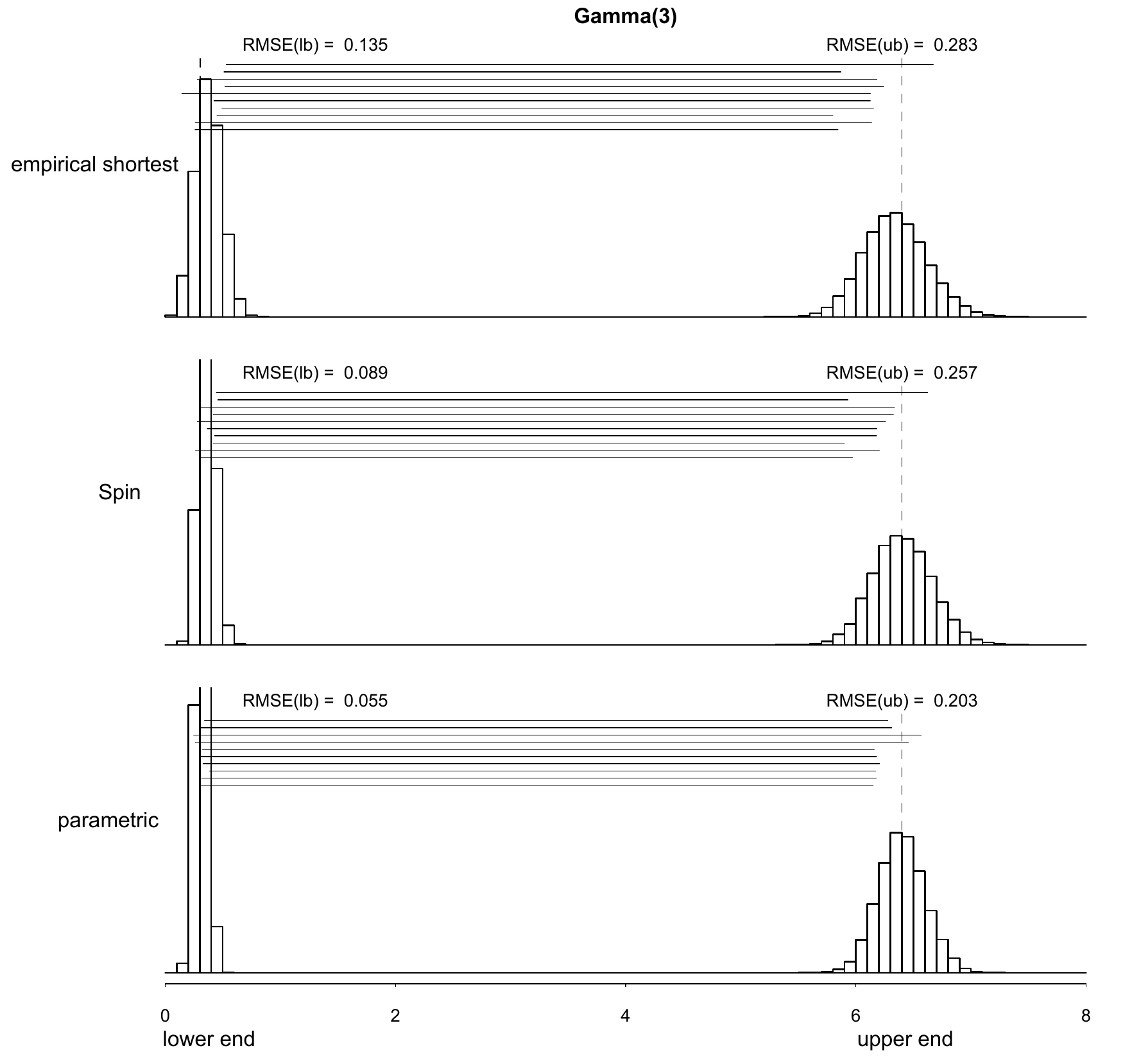}
\vspace{.2cm}
\caption{\em Spin for an asymmetric distribution.  95\% intervals for the gamma distributions with shape parameter 3, as estimated from 500 independent draws. Each horizontal bar
  represents an interval from one simulation. The histograms are based
  on results from 20,000 simulations. The dotted vertical lines
  represent the true endpoints of the HPD interval.  Spin outperforms the empirical shortest interval.  The interval obtained from a parametric fit is even better but this approach cannot be applied in general; rather, it represents an optimality bound for any method.}
\label{gamma}
\end{figure}

Figure \ref{gamma} shows the empirical shortest, Spin, and parametric
intervals constructed from 500 samples of the gamma distribution with shape parameter 3. Spin gets more accurate endpoint
estimates than empirical shortest intervals. Specifically, for the lower end where the density is relatively
high, Spin estimates are less variable, and for the upper end
at the tail of the density, Spin shows a smaller bias. Again, the lowest RMSE comes from taking advantage of
the parametric form of the posterior distribution, which is rarely practical in real
MCMC applications.  Hence the RMSE using the parametric form represents
a rough lower bound on the  Monte Carlo error in any HPD computed from
simulations.

\begin{figure}
\centering
\includegraphics[width=\textwidth]{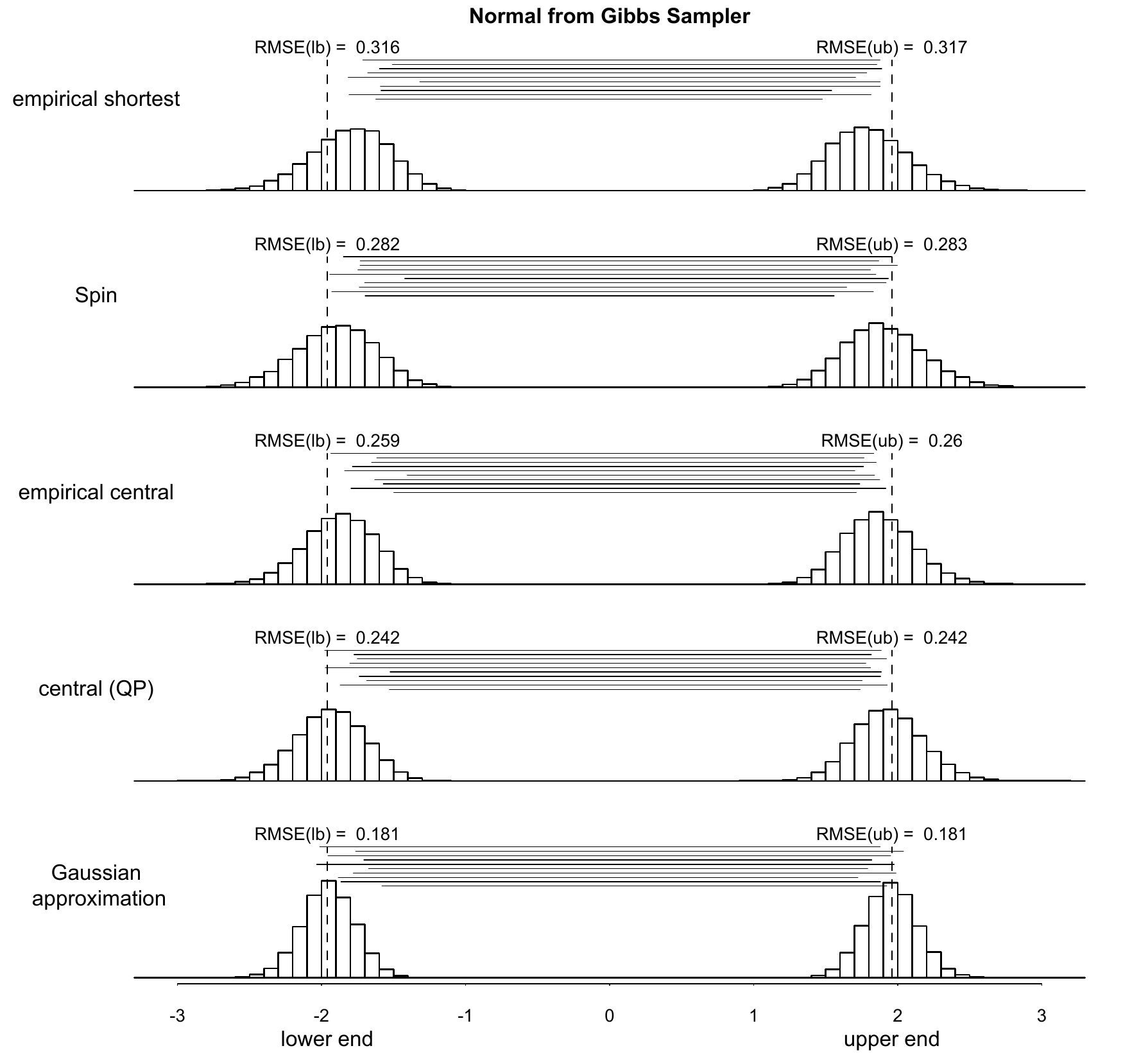}
\vspace{.2cm}
\caption{\em Spin for MCMC samples.  95\% intervals for normal samples
  from Gibbs sampler, in each case based on 200 draws. Each horizontal bar
  represents an interval from one simulation. The histograms are based
  on results from 20,000 simulations. The dotted vertical lines
  represent the true endpoints of the HPD intervals.  Spin greatly
  outperforms the raw empirical shortest interval.  The central
  interval (and its quadratic programming improvement) does even
  better.  Again the intervals estimated by fitting a Gaussian distribution do the best.}
\label{gibbs}
\end{figure}

Figure \ref{gibbs} shows the intervals constructed for MCMC normal
samples. Specifically, the Gibbs sampler is used to draw samples from a
standard bivariate normal distribution with correlation 0.9. We use this example to explore how Spin works on simulations with high autocorrelation. Two chains each with
1000 samples are drawn with Gibbs sampling. For one variable, every ten draws are recorded for Spin
construction, resulting in 200 samples, which is roughly the level of the
effective sample size in this case. This is a typical senario in
practice when MCMC techniques are adopted for multivariate distributions. Again Spin greatly outperforms the
empirical shortest interval in case of highly correlated draws.

\begin{figure}
\centering
\includegraphics[width=\textwidth]{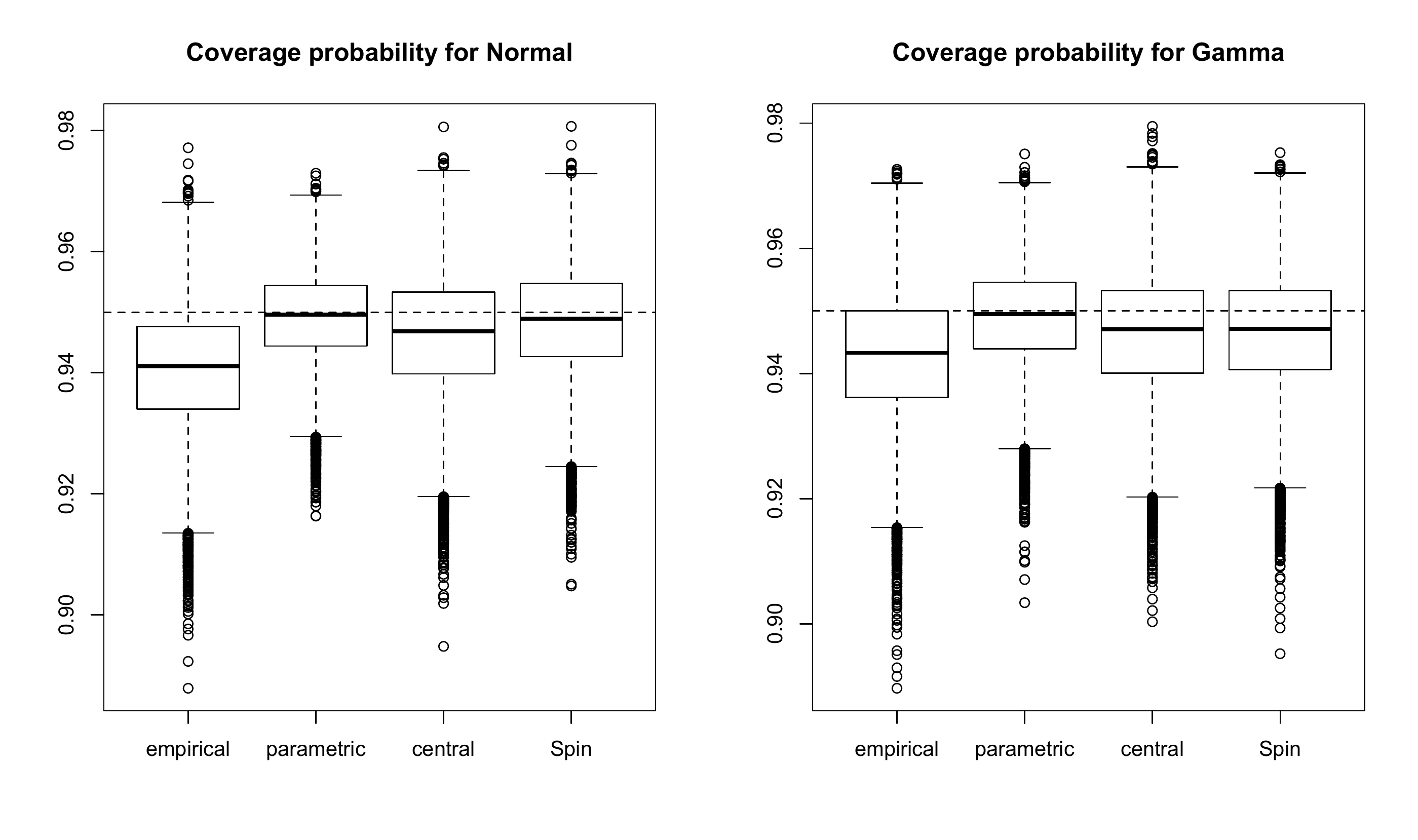}
\caption{\em Distribution of coverage probabilities for Spin and other 95\% intervals calculated based on 500 simulations for the normal and gamma(3) distributions.}
\label{cp}
\end{figure}

We further investigate coverage probabilities of the different
intervals constructed (Figure \ref{cp}).
Empirical shortest intervals have the lowest coverage probability,
which is as expected since they are biased towards shorter intervals (see Figure
\ref{sym} and Figure \ref{gamma}). Coverage probabilities of Spin are closer to the nominal coverage (95\%) for both normal and
gamma distributions. Comparable coverage is observed for central
intervals. As expected, parametric intervals represent a gold standard and have the most accurate coverage.

\begin{figure}
\centering
\includegraphics[width=.5\textwidth]{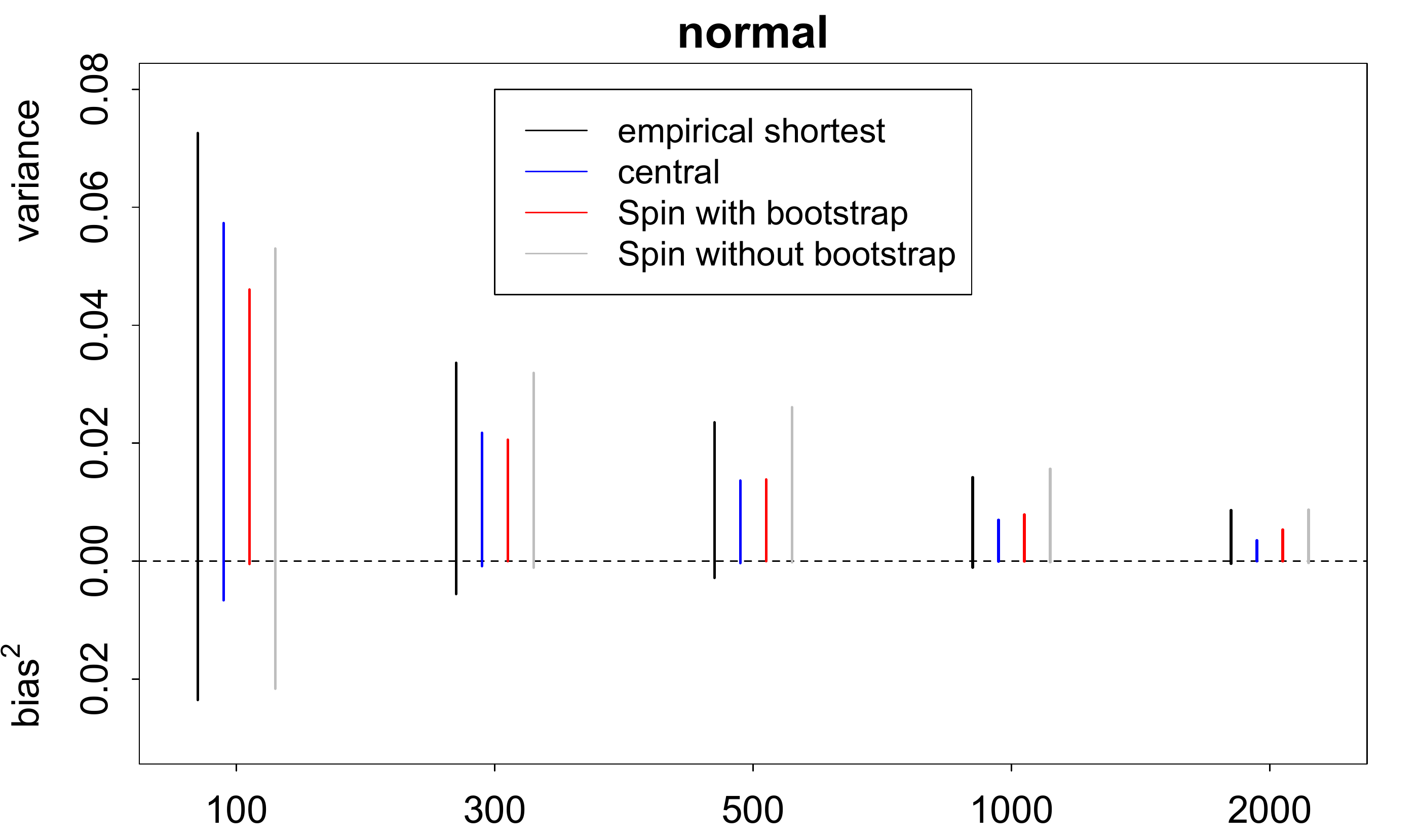}\includegraphics[width=.5\textwidth]{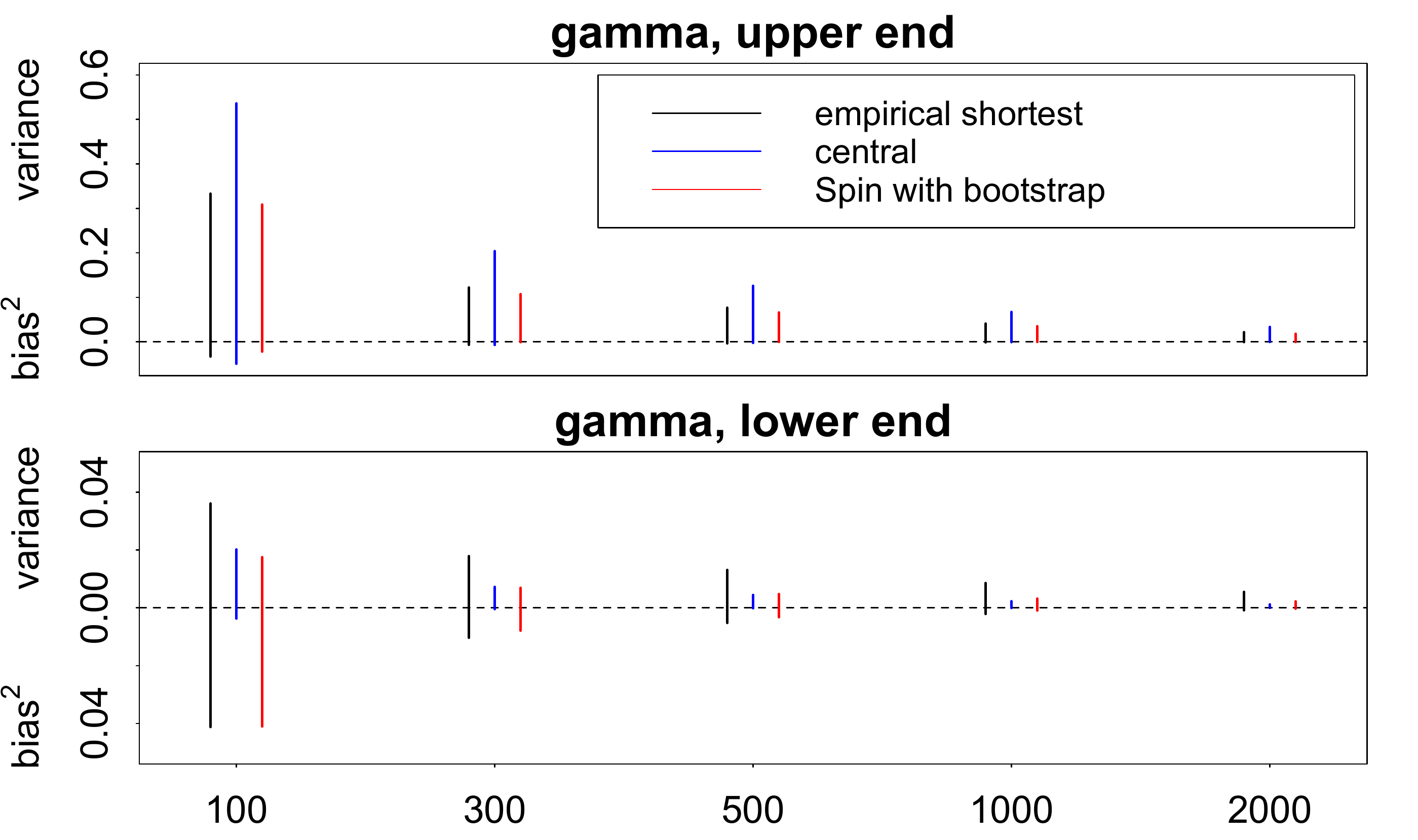}
\vspace{.5cm}
\caption{\em Bias-variance decomposition for 95\% intervals for normal and
  gamma(3) examples, as a function of the number of simulation draws.
  Because of the symmetry of the normal distribution, we averaged its
  errors for upper and lower endpoints. Results from Spin without
  bootstrap are shown for normal for description purpose.}
\label{bvgamma}
\end{figure}

Figure \ref{bvgamma} shows the bias-variance
decomposition of different interval estimates for normal and gamma distributions under sample sizes
100, 300, 500, 1,000 and 2,000. We average lower and upper ends for the normal case
due to symmetry. For both distributions, Spin has both well-reduced variance
and bias compared to the empirical shortest intervals. The upper end estimates of
empirical central intervals for the gamma have a large variance since the
corresponding density is low so the observed simulations in this
region are more variable. It is worth pointing out that the
computational time for Spin is negligible compared to sampling, thus it
is a more efficient way to obtain improved interval estimates. In the
normal example shown in the left panel in Figure \ref{bvgamma}, rather
than increasing the sample size from 300 to 500 to reduce error, one can spend less time to compute Spin with the 300
samples and get a even better interval.

We also carried out experiments with even bigger samples
and intervals of other coverages (90\% and 50\%), and got
similar results.  Spin beats the empirical shortest interval in RMSE (which makes sense, given that Spin is optimizing over a class of estimators that includes the empirical shortest as a special case).

\section{Results for two real-data examples}\label{res2}

In this section, we apply our methods to two applied examples of hierarchical Bayesian models, one from education and one from sociology. In the first example, we
show the advantages of Spin over central and empirical shortest intervals;
in the second example, we demonstrate the routine use of Spin to summarize posterior inferences.

\begin{figure}
\centering
\includegraphics[width=\textwidth]{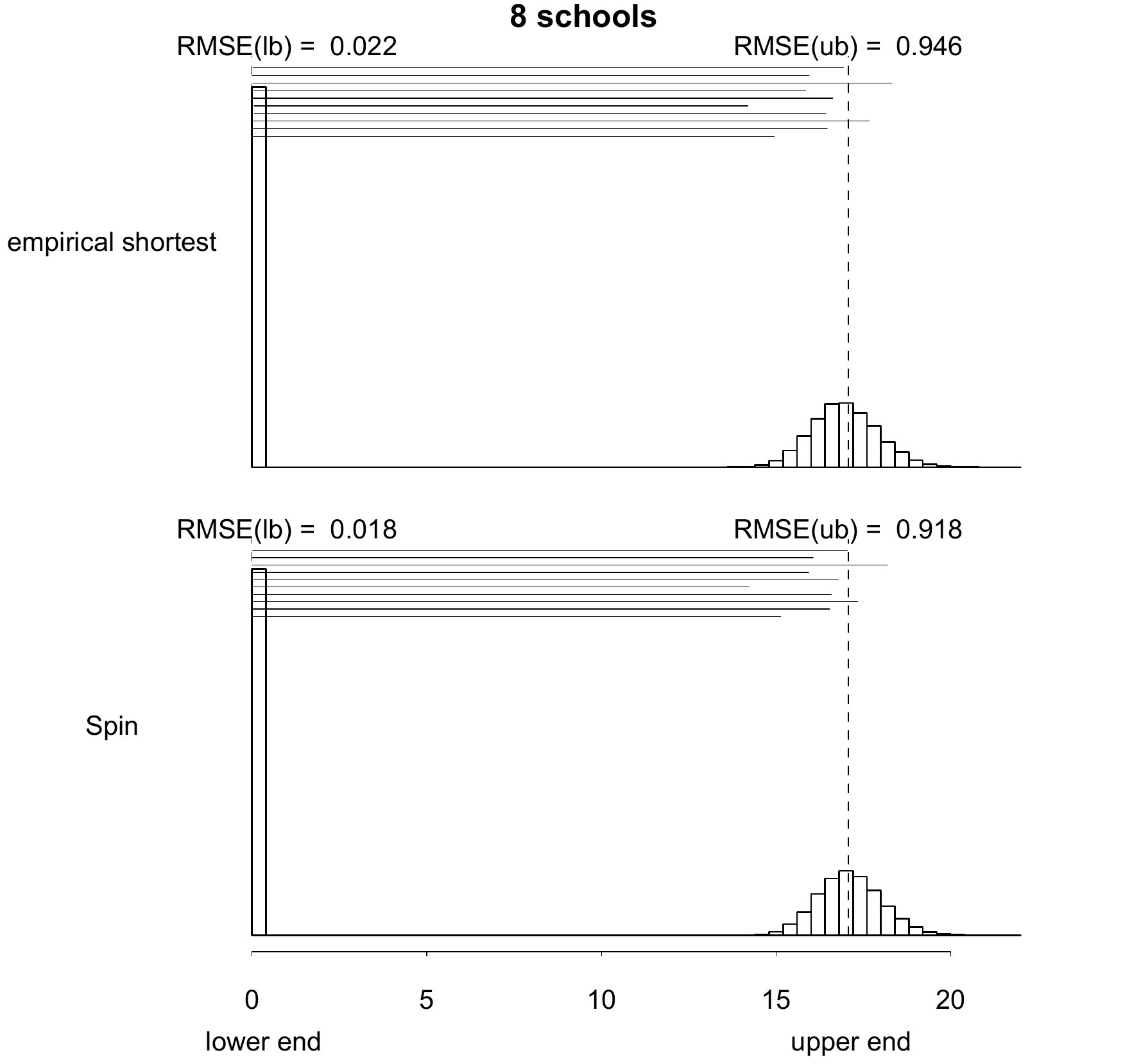}
\vspace{.2cm}
\caption{\em Spin for the group-level standard deviation parameter in the eight schools example, as estimated from 500 independent draws from the posterior distribution (which is the right density curve in Figure \ref{fig1}, a distribution that is constrained to be nonnegative and has a minimum at zero). The histograms in this figure are based on results from 20,000 simulations. The dotted vertical lines represent the true endpoints of the HPD interval as calculated numerically from the posterior density.  Spin does better than the empirical shortest interval, especially at the left end, where its smoothing tends to (correctly) pull the lower bound of the interval all the way to the boundary at 0.}
\label{8sch}
\end{figure}

\begin{figure}
\centering
\includegraphics[width=\textwidth]{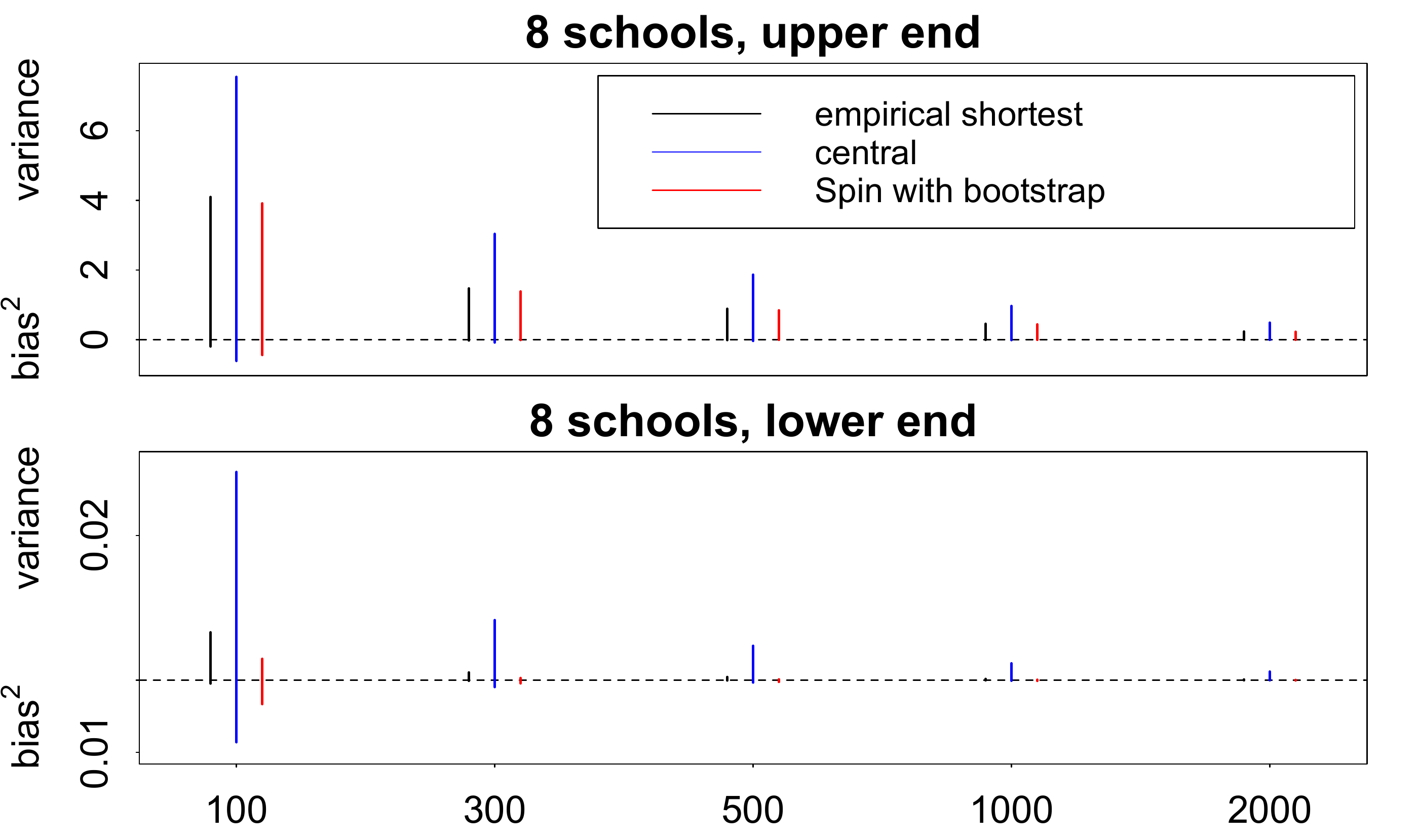}
\vspace{.2cm}
\caption{\em Bias-variance decomposition for 95\% intervals for the eight-school example, as a function of the number of simulation draws.}
\label{8bv}
\end{figure}

Our first example is a Bayesian analysis from Rubin (1980) of a hierarchical model of data from a set of experiments performed on eight schools.  The group-level scale parameter (which corresponds to the between-school standard deviation of the underlying treatment effects) has a posterior distribution that is asymmetric with a mode at zero (as shown in the right panel of Figure
\ref{fig1}).  Central probability intervals for this scale parameter (as presented, for example, in the analysis of these data by Gelman et al., 1995) are unsatisfying in that they exclude a small segment near zero where the posterior distribution is in fact largest.  Figure
\ref{8sch} shows the 95\% empirical shortest intervals and Spin constructed
from 500 draws. The results of empirical shortest intervals for 8
schools are from including the zero point in the simulations. Spin has smaller RMSE than
both empirical shortest and central intervals (Figure \ref{8sch} and
Figure \ref{8bv}).

\begin{figure}
\centering
\includegraphics[width=120mm]{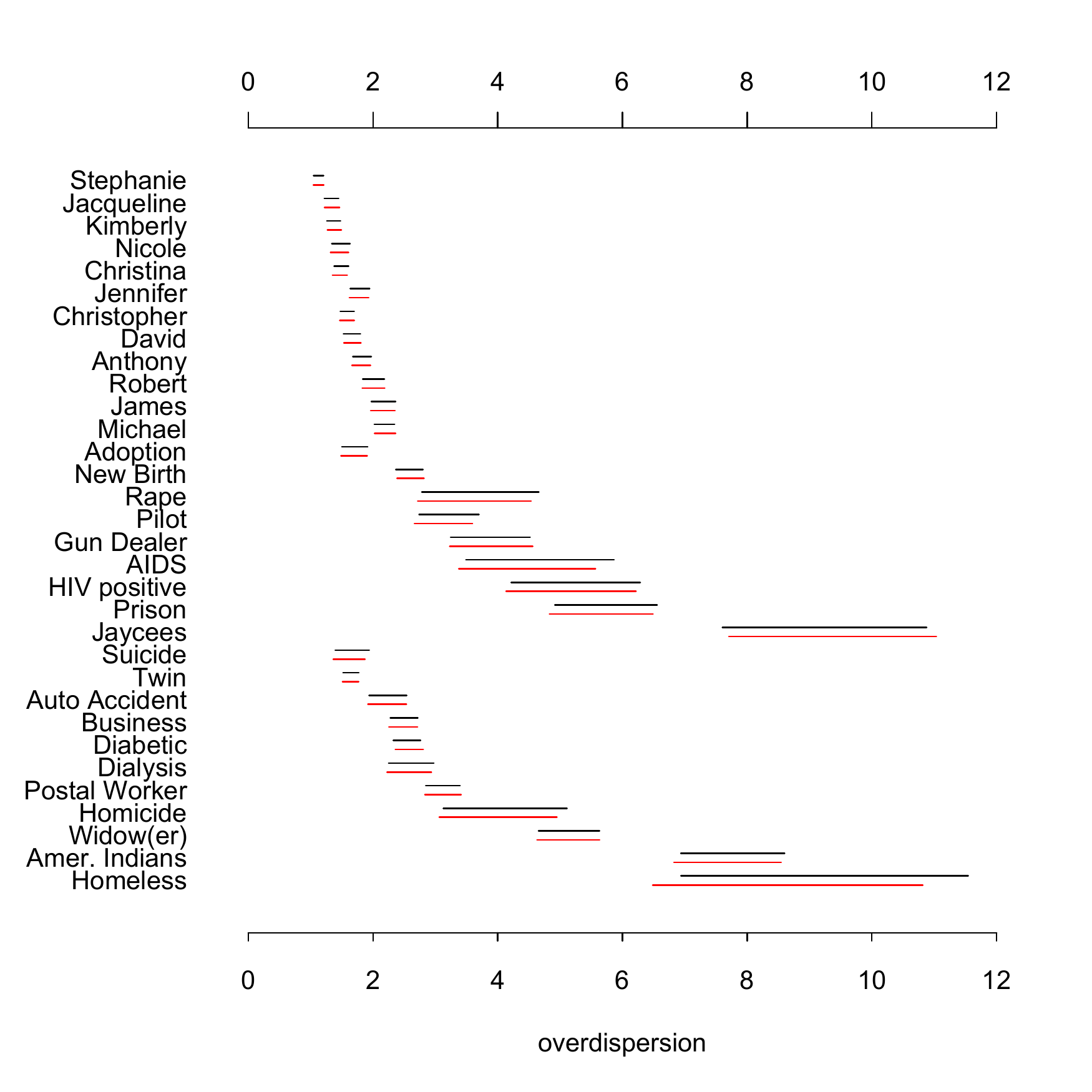}
\caption{\em 95\% central intervals (black lines) and Spins (red lines) for the overdispersion parameters in the ``How many X's do you know?'' study.  The parameter in each row is a measure of the social clustering of a certain group in the general population:  groups of people identified by first names have low overdispersion and are close to randomly distributed in the social network, whereas categories such as airline pilots or American Indians are more overdispersed (that is, non-randomly distributed).  We prefer the Spins as providing better summaries of these highly skewed posterior distributions.  However, the differences between central intervals and Spins are not large; our real point here is not that the Spins are much better but that they will work just fine in routine applied Bayesian practice, satisfying the same needs as were served by central intervals but without that annoying behavior when distributions are highly asymmetric.}
\label{overdis}
\end{figure}

For our final example, we fit the social network model of Zheng et al.\ (2006) using MCMC and construct 95\% Spins for the overdispersion parameters based on 200 posterior draws. The posterior is
asymmetric and bounded below at 1. Figure \ref{overdis}
is a partial replot of Figure 4 from  Zheng et al.\ (2006) with Spins
added. For this type of asymmetric posterior we prefer the estimated HPDs to the corresponding central intervals as HPDs more precisely capture the values of the parameter that are supported by the posterior distribution.

\section{Discussion}\label{discussion}

We have presented a novel optimal approach for
constructing reduced-error shortest probability intervals
(Spin). Simulation studies and real data examples show the advantage
of Spin over the empirical shortest interval. Another commonly used
interval estimate in Bayesian inference is the central interval. For
symmetric distributions, central intervals and HPDs are the same;
otherwise we agree with Box and Tiao (1973) that the HPD is generally preferable to the central
interval as an inferential summary (Figure \ref{fig1}). In our
examples we have found that for symmetric distributions Spin and
empirical central intervals have comparable RMSEs and coverage
probabilities (Figures \ref{sym}, \ref{cp}, and \ref{bvgamma}). Therefore we recommend Spin as a default procedure for computing HPD intervals from simulations, as it is as computationally stable as the central intervals which are currently standard in practice.

We set the bandwidth parameter  $b$ in (\ref{c}) to $\sqrt{n}$, which seems to work well for a variety of distributions. We also carried out sensitivity analysis by varying $b$ and found that large $b$ tends to result in more stable endpoint estimates where the density is relatively high but can lead to noisy estimates where the density is low.  This makes sense:  in low-density regions, adding more points to the weighted average may introduce noise instead of true signals. Based on our experiments, we believe the default value $b=\sqrt{n}$ is a safe general choice.

Our approach can be considered more generally as a method of using weighted averages of order statistics to construct optimal interval estimates. One can replace $Q(\Delta^*)$ in (\ref{MSE}) by the endpoints of any reasonable empirical interval estimates, and obtain improved intervals by using our quadratic programming strategy (such as the improved central intervals shown in Figure \ref{sym}).

One concern that arises is the computational cost of performing Spin itself.  Our simulations show Spin intervals to have better simulation coverage and appreciably lower mean squared error compared to the empirical HPD, but for simple problems in which one can quickly draw direct posterior simulations, it could be simpler to forget Spin and instead just double  the size of the posterior sample.  More and more, though, we find ourselves computing Bayesian models using elaborate Markov chain simulation algorithms for which it can take hundreds of thousands of steps, and hours or even days of computing time, to obtain an effective sample size of a few hundred posterior simulation draws.  In this case, the Spin calculations are a small price to pay for obtaining more accurate and stable HPD intervals.

We have demonstrated that our Spin procedure works well in a range of
theoretical and applied problems, that it is simulation-consistent,
computationally feasible, addresses the boundary problem, and is
optimal within a certain class of procedures that include the
empirical shortest interval as a special case.  We do not claim,
however, that the procedure is optimal in any universal sense.  We see
the key contribution of the present paper as developing a practical
procedure to compute shortest probability intervals from simulation in
a way that is superior to the naive approach and is competitive (in
terms of simulation variability) with central probability intervals.
Now that Spin can be computed routinely, we anticipate further
research improvements on posterior summaries.


\section*{REFERENCES}

\noindent
\bibitem Box, G. E. P., and Tiao, G. C. (1973).  {\em Bayesian Inference in Statistical Analysis}.
New York:  Wiley Classics.

\bibitem Chen, M. H., and Shao, Q. M. (1999).
Monte Carlo estimation of Bayesian credible and HPD intervals.
{\em  Journal of Computational and Graphical Statistics} {\bf 8}, 69--92.

\bibitem David, H. A., and Nagaraja, H. N. (2003).
{\em Order Statistics}, third edition.
New York:  Wiley.

\bibitem Efron, B. (1979).  Bootstrap methods: Another look at the jackknife. {\em Annals of Statistics} {\bf 7}, 1-–26.

\bibitem Gelman, A., Carlin, J. B., Stern, H. S., and Rubin, D. B. (1995).
{\em Bayesian Data Analysis}.  London:  CRC Press.

\bibitem Gelman, A., and Shirley, K. (2011).  Inference from simulations and monitoring convergence. In {\em Handbook of Markov Chain Monte Carlo}, ed.\ S. Brooks, A. Gelman, G. Jones, and X. L. Meng, 163--174. London: CRC Press. 

\bibitem R Development Core Team (2011).
{\em R: A Language and Environment for Statistical Computing}.
Vienna, Austria:  R Foundation for Statistical Computing.

\bibitem Rubin, D. B. (1981).  Estimation in parallel randomized experiments.
{\em Journal of Educational Statistics} {\bf 6}, 377--401.

\bibitem Spiegelhalter, D., Thomas, A., Best, N., Gilks, W., and Lunn, D.
(1994, 2002).
BUGS:  Bayesian inference using Gibbs sampling.  MRC Biostatistics Unit, Cambridge, England.
{\tt www.mrc-bsu.cam.ac.uk/bugs/}

\bibitem van der Vaart, A. W. (1998).  {\em Asymptotic Statistics}.
Cambridge University Press.

\bibitem Zheng, T., Salganik, M. J., and Gelman, A. (2006).
How many people do you know in prison?: Using overdispersion in count data to estimate social structure in networks.
{\em Journal of the American Statistical Association} {\bf 101}, 409--423.

\end{document}